# Source Code Protection for Applications Written in Microsoft Excel and Google Spreadsheet

Thomas A. Grossman
University of San Francisco, School of Business & Management, San Francisco CA  94117
tagrossman@usfca.edu

**ABSTRACT**

Spreadsheets are used to develop application software that is distributed to users. Unfortunately, the users often have the ability to change the programming statements ("source code") of the spreadsheet application. This causes a host of problems. By critically examining the suitability of spreadsheet computer programming languages for application development, six "application development features" are identified, with source code protection being the most important. We investigate the status of these features and discuss how they might be implemented in the dominant Microsoft Excel spreadsheet and in the new Google Spreadsheet. Although Google Spreadsheet currently provides no source code control, its web-centric delivery model offers technical advantages for future provision of a rich set of features. Excel has a number of tools that can be combined to provide "pretty good protection" of source code, but weak passwords reduce its robustness. User access to Excel source code must be considered a programmer choice rather than an attribute of the spreadsheet.

1.  **INTRODUCTION**

Spreadsheets can be a powerful language to write application software for distribution to users. Unfortunately, the users often have access to the programming statements (programmers call these the "source code") of the spreadsheet application. This makes it easy for users to change the application they are using, while they are using it. Although this feature is desirable for exploratory modeling, collaborative work, and rapid development, it plays havoc with attempts to insure accuracy, prevent rampant user customizations, provide version control, have a workable update and maintenance process, and institute effective information system controls. Someone ought to do something.

This paper discusses what might be done and who can do it. Spreadsheets (there are over a dozen of them now, in two distinct classes) should be thought of as a class of computer programming languages, and critically evaluated for their suitability for application development. This paper proposes six features that the designers of spreadsheet computer programming languages should include in their products, and explores how they are handled—and might be implemented—in the dominant Microsoft Excel spreadsheet and in the "software-as-a-service" Google Spreadsheet. Software-as-a-Service spreadsheets seem to have a technical advantage in delivering these features. The paper also explains how to provide "pretty good protection" of source code in the dominant Excel spreadsheet.

If the six application development features were implemented in spreadsheet computer programming languages, it could revolutionize the development of application software in spreadsheets. This has potential to substantially reduce the risks and problems associated with spreadsheets, reduce lifecycle costs, allow effective information systems controls, and make spreadsheets suitable for high-control uses such as Sarbanes-Oxley compliance.

1.1.  **Literature Survey**

I am unaware of any literature that discusses creating spreadsheet applications, or explores in any detail the points raised in this paper. [Bewig 2005] mentions users as distinct from developers, but focuses on what one might call 'power developer' skills. He touches on many source code protection concepts in Excel. [Powell & Baker 2007 pp. 106-108] have a short section on designing a workbook as a decision support system used by multiple people, and discuss some Excel protection features. [O'Beirne 2005] discusses passwords and their cracking. [Read and Batson 1999] mention cell protection and Excel data validation.

## 1.2. Structure of Paper

The paper begins (section 2) with a discussion of spreadsheet applications, and a proposed categorization into "intentional spreadsheet applications", "*de facto* spreadsheet applications" and "accidental spreadsheet legacy applications". This leads into section 3 which argues that spreadsheets should be thought of as a class of computer programming languages for writing application software. Computer scientists routinely evaluate computer programming languages for their suitability for various tasks, and section 4 of the paper concludes that spreadsheet languages would be much better suited for application programming if they prevented users from accessing the source code. Section 5 presents six application development features that are desirable in a spreadsheet computer programming language.

Section 6 discusses a new class of spreadsheet languages delivered over the web, and uses Google Spreadsheet as an example. Section 7 discusses the excellent prospects for these "software-as-a-service" spreadsheets to provide application development features, and sketches how these might be implemented for the case of Google Spreadsheet by adapting the existing sharing feature. Section 8 explores "local" spreadsheets, focusing on Microsoft Excel, and discusses Excel's tolerably effective source code protection features, which are hindered by poor password encryption. Section 9 summarizes Excel protections and concludes that user access to source code is not an attribute of Excel but rather is a choice by the programmer. Section 10 discusses how to provide application development features in local spreadsheets. Conclusions and further research are discussed in section 11.

## 2. MOTIVATION: SPREADSHEET APPLICATION SOFTWARE

Application software is "programs written for a specific application to perform functions specified by end users" [Laudon and Laudon 2006]. Spreadsheet application software is a computer program written in a spreadsheet and provided to end users. I propose three categories for classifying spreadsheet application software:

**Intentional spreadsheet applications** are software developed in a spreadsheet and purposefully deployed to users. [Grossman, Mehrotra, and Özlük 2006] give three examples built by professional developers. These are a spreadsheet application for sale (dozens of worksheets in three workbooks totaling 70 MB); a spreadsheet application distributed to thousands of users in a multi-national company; and a spreadsheet decision support system provided by a consulting firm to a client organization. These spreadsheets are very much the product of traditional application development, where the developers selected the spreadsheet as the appropriate programming language. Intentional spreadsheet applications are undoubtedly constructed by casual programmers as well as professional developers.

**De facto spreadsheet applications** are spreadsheets constructed for personal use that are distributed to other people. There is widespread awareness of spreadsheets that are built for individual use that find their way onto other people's computers. These spreadsheets can take on a life of their own, and present many well-known challenges such as risk of error, improper usage, inefficient updating and maintenance, inappropriate and uncontrolled modification, and so forth. These spreadsheets were not constructed with the intention of use by other people, but become applications by dint of distribution to users.

**Spreadsheet accidental legacy applications** are spreadsheets inherited by a person when they start a new job. [Grossman, Mehrotra, and Özlük 2006] present three examples, including one that led to a multi-million dollar financial loss. Industry is recognizing spreadsheet legacy applications; [Gardner 2006] tells us that "a new business [of spreadsheet management] is being built around the millions of old spreadsheet applications that were developed months and years ago, but are still present in companies' financial records".

3. **THE SPREADSHEET IS A COMPUTER PROGRAMMING LANGUAGE FOR WRITING APPLICATIONS**

Spreadsheets are being used for application development, whether by professional or amateur developers, acting with intention or by accident. However, spreadsheets receive little attention and (sometimes little respect) for their use in this role. I believe that spreadsheets merit careful study as a powerful application development language.

**3.1. Not an Application, but a Language for Developing Applications**

The spreadsheet is commonly referred to as an "application", akin to a word processor or an email client. I think it is a mistake to refer to the spreadsheet as an "application". The spreadsheet should be considered a class of computer programming languages. Like most computer programming languages, a spreadsheet can be used to write applications that are delivered to users, as well as for personal productivity and other uses.

To be considered a class of computer programming language, there must be substantial choice, and there is. Wikipedia tells us there are over a dozen spreadsheet computer programming languages available. The dominant spreadsheet language is Microsoft Excel, which uses the traditional "local" software delivery model, where the spreadsheet product is installed on the user's hard drive. Other local spreadsheet languages include OpenOffice Calc, Lotus 1-2-3, Quattro Pro, KOffice KSpread, and Gnumeric.

A new delivery model called "Software as a Service" (SaaS) runs a spreadsheet computer programming language on a remote server. Programmers and users interact with the spreadsheet language using a web browser. Google Spreadsheet gets the most press, but others include EditGrid, iRows, Simple Spreadsheet, ThinkFree Calc, wikiCalc and Zoho Sheet.

Because SaaS spreadsheets reside on a remote server rather than on the user's hard drive, in principle the programmer has a high degree of control over who can access the spreadsheet and what they can see. A SaaS spreadsheet language may have advantages for restricting what user activities with the spreadsheet.

**3.2. Strengths and Weaknesses of Spreadsheet Computer Programming Languages**

Spreadsheet computer programming languages have many attractive programming features. Most prominent is a beguiling accessibility that allows people who are not professional programmers to quickly write software with little or no training. Spreadsheets are an excellent rapid development environment, and support collaborative interactions. Spreadsheet languages are attractive for application development because users are familiar with the interface and deployment is easy due to the ubiquity of Excel. Spreadsheet languages are attractive to users because they are easily extended and connected to user-built spreadsheets, although these features are a mixed blessing.

As with any class of programming languages, there are drawbacks. Errors are a concern, as they are in all software, and some question the wisdom of letting non-professionals program. The most significant drawback of a spreadsheet from the perspective of application programming is that in general the user can modify the source code. What this means is that the user can access and change the computer programming statements of the application that he is using, while he is using it!

## 4. THE PRIZE: A SPREADSHEET COMPUTER PROGRAMMING LANGUAGE THAT IS GENUINELY SUITABLE FOR APPLICATION PROGRAMMING

Traditional programming languages prohibit or at least make it difficult for the user to access or modify the source code. Compiled languages such as C++ provide an executable but not the source code, and the source code cannot easily be reverse engineered from the executable.

In contrast, spreadsheets make it easy for the user to mess about with the source code. The user receives not an executable, but a copy of the source code, along with an integrated user-friendly source code editor, non-cryptic programming statements, and an instantaneous code interpreter. It's virtually an invitation to the user to modify the source code. Because other interpreted languages don't provide this bundle of features, they are less vulnerable.

The spreadsheet application programmer faces unique challenges. It is easy for someone who receives a spreadsheet application to modify the underlying programming statements, by intention or by accident. This can lead to errors. It confuses any versioning structure the programmer might have, and greatly hinders maintenance and adoption of new versions. It makes a mockery of any information system controls and is a root cause of "the subversive spreadsheet". For financial organizations it is inconsistent with Sarbanes-Oxley compliance.

There would be great benefit if spreadsheet application programmers could enjoy the same level of control that, say, C++ application programmers have in the distribution and control of their source code. This would allow developers to use spreadsheet languages with much greater confidence, and permit their use in high-security environments such as Sarbanes-Oxley. This would be a great prize. The question now become, what features are needed to obtain this prize, and which are available in existing spreadsheets?

## 5. THE CHALLENGE: SPREADSHEET LANGUAGE FEATURES TO SUPPORT APPLICATION DEVELOPMENT

To support the use of the spreadsheet computer programming language for writing applications, the publishers of spreadsheet languages need to provide certain features. The ideal approach would be to consider "application development features" during the design phase of the spreadsheet language. Failing this, the task becomes adding features to an existing language. There are six features, with only the first being absolutely necessary.

The essential feature is to **(1) protect source code from user changes.** This prevents users from accessing and modifying the source code. One way to think about this is to provide the ability to designate each cell as "*full-access*" or "*display-access*". The user has full privileges over *full-access* cells, including the ability to view and edit cell contents. The user has minimal privileges over *display-access* cells, and can see displayed values but not cell contents, including when the spreadsheet is saved to a local hard drive. An enhanced feature would allow cells to be designated "*no-access*" where the user can't see them. One can easily imagine other possibilities and extensions, but these are the key elements.

It is obviously desirable to **(2) prevent user customization** by prohibiting adding programming statements to the spreadsheet. This enhances application integrity and is essential for version control. More subtly, it is important to consider the possibility of linking a user spreadsheet into the application spreadsheet to create a local system; this might be a beneficial integration or a problematic rogue system. Therefore, the programmer should have the ability to **(3) restrict linking** user spreadsheets into the application. Finally, it would be helpful to **(4) prevent users from sharing the application** without the developer's permission. This would solve the problem of proliferation.

It is desirable to **(5) provide easy control of protection features** using a well-designed user interface where the programmer can select the features he wants to use. Finally, it is important to **(6) prevent user tampering with protection features** by employing effective password protection, and avoiding simple circumventions such as the Excel copy-and-paste evasion discussed below.

## 6. SaaS SPREADSHEETS: THE CASE OF GOOGLE SPREADSHEET

A SaaS spreadsheet is resident on a server, and information is sent to the user's computer only as appropriate, whereas with a local spreadsheet all information is present on the user's computer. For this reason, SaaS spreadsheets seem advantageous for source code control, and it is argued that SaaS systems can provide excellent security [Wikipedia 2007]. SaaS security can be based on user accounts authenticated with log-in passwords over SSL connections, which extensive e-commerce experience shows to be very effective.

Since SaaS spreadsheets are new, I will use Google Spreadsheet (docs.google.com) as an example. The capabilities of the Google Spreadsheet beta release pale in comparison to Excel. However, Google spreadsheet is not intended to be a copy of Excel on the internet; instead it provides innovative "web-centric" features, including simultaneous editing by two users and API "hooks" to connect the spreadsheet to other applications.

Google Spreadsheets controls access through log-in passwords. It has a sharing feature that a programmer can use to grant access to other people. A "*collaborator*" is granted full control of the spreadsheet. A "*viewer*" is granted the ability to see spreadsheet but not edit it. A *viewer* cannot change nor copy cell contents. If the a viewer copies a range of cells into Excel, only the values are copied, not the cell formulas. A *viewer* can observe the complete cell contents (including formulas) as shown in the screenshot below, where cell A3 is partially displayed but its contents are observable by hovering the cursor over a field in the bottom bar:

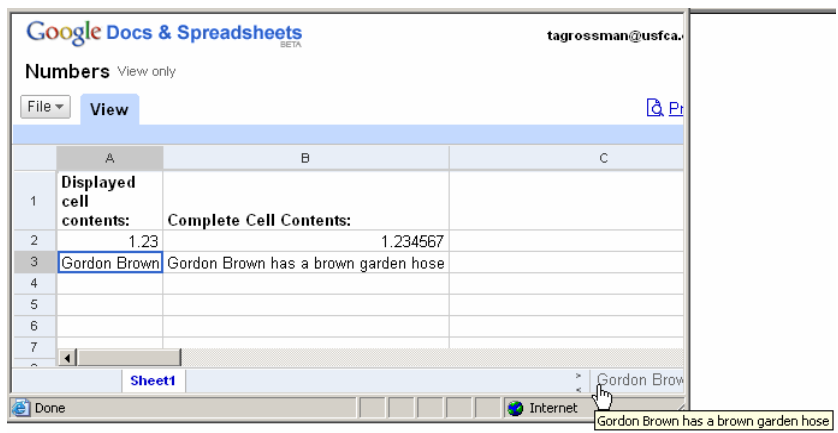

At the time of writing, information in the spreadsheet can not be hidden, and the *viewer* can always see the source code.

## 7. FEATURES REQUIRED IN SaaS SPREADSHEETS TO SUPPORT APPLICATION DEVELOPMENT: THE CASE OF GOOGLE SPREADSHEET

If the source code were protected, a SaaS spreadsheet could provide significant advantages for writing and distributing applications. The programmer could maintain on a server a master copy of the application that large numbers of users access remotely. A new version could be rolled out at the click of a button, with the old version safely locked away.

In Google Spreadsheet, a "sharing" feature controls the information that is sent to the user. In principle, Google Spreadsheet could provide an excellent facility for protecting source code (feature 1 in section 5) by creating a new sharing class, which I'll call "*limited-user*". In its simplest form, a *limited-user* would have the following privileges:

1. See all values displayed in the spreadsheet.

2. For cells containing data (text or numbers): the access control state is *full-access*. The user can observe complete cell contents and edit cell contents; the bottom bar detail is enabled.
3. For cells containing cell formulas: the access control state is *display-access*. The user can not observe cell contents nor edit the cell; double-clicking the cell and the bottom bar detail are disabled.
4. A copy of the spreadsheet saved locally contains values not cell contents.

A slightly more sophisticated approach would allow the programmer to specify data cells as full-access. This would allow the programmer to safeguard certain text and numeric values (such as labels and model parameters) while permitting the user to change input values.

Google's invitation-only sharing system prevents users from sharing the application (feature 4). Google's log-in password system seems suitable for preventing user tampering (feature 6) but ideally should be implemented using SSL. A *limited-user* could be prevented from writing cell formulas (feature 2) and linking to the spreadsheet (feature 3). It seems likely that a skilled user interface designer could allow easy control (feature 5).

It appears that a SaaS spreadsheet programming language could deploy source code protection that users could not evade, and in addition provide all six of the desired application programming features.

**8. LOCAL SPREADSHEETS: THE COMPLEX CASE OF MICROSOFT EXCEL**

Because of its ubiquity, Excel merits special attention. Excel has several features that can be used together to safeguard source code. Because the safeguards are unduly complex and some combinations of settings are insecure, it is worth reviewing them. Menu sequences in Excel 2003 are shown in normal font. Menu sequences in Excel 2007 are shown in *italics*.

**8.1. Cell Protection Formats: Lock and Hide Cells**

The Cell Protection Formats control whether a user can see or edit the source code in a cell. Cell Protection Formats are controlled in the Protection Tab of the Format Cells dialog box, accessed via Format\Cells…\Protection\ or *Home\Cells\Format\Cells\Protection\*. The Cell Protection formats are activated only when Sheet Protection (section 8.2) is enabled.

If a cell is formatted Locked, the user can not edit its source code; if unLocked the user can edit its source code. If the cell is formatted Hidden, the source code in a cell can not be seen (even in Formula View); if unHidden the source code can be seen.

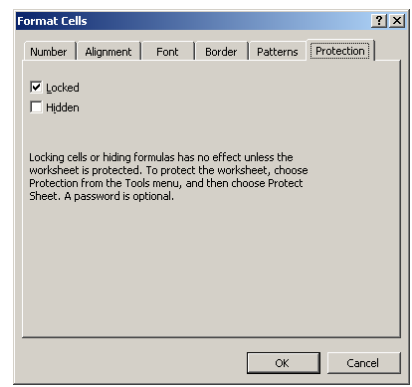

When Sheet Protection is disabled, the Cell Protection Formats are turned off and all cells are unLocked and unHidden. This is the usual Excel behavior, where the user has the keys to the kingdom and can see and edit all source code.

The default cell format is Locked and unHidden, so when Sheet Protection is enabled, the standard behavior is that cell contents can be seen but not changed. Formatting a cell as unLocked and Hidden is not recommended, because it allows the user to edit a cell whose contents he can not see. Strongest control is obtained by formatting a cell as Locked and Hidden, which prevents the user from seeing or editing the source code.

**8.2. Sheet Protection: Control Cell Selection and Activate Cell Protection Formats**

Sheet Protection has two types of effects. First, it controls whether the Cell Protection Formats (section 8.1) are enabled. Second, it controls the ability of the user to select cells; format cells (including conditional formatting); insert and delete rows, columns and hyperlinks; use analytical tools; and edit objects and scenarios (plus it disables the precedence/dependence arrows on the formula auditing toolbar, which is not programmer-selectable). Sheet Protection is controlled in the Protect Sheet dialog box, accessed via Tools\Protect Sheet\or *Review\Changes\Protect Sheet\*.

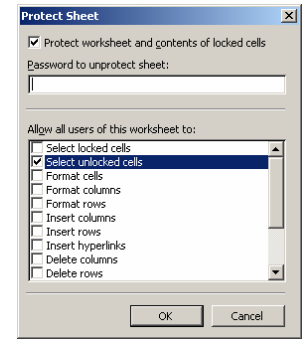

When Sheet Protection is enabled, the default settings are "Select locked cells" and "Select unlocked cells". Disabling "Select locked cells" renders it impossible for the user to select a cell that he can not change (due to its being Locked), which reduces user frustration. There is a more subtle benefit as well. When the user navigates in the spreadsheet using the Enter key, Tab key, or Arrow keys, Excel cycles only through the unLocked cells. This is equivalent to a traditional application where the user can move the cursor only to data entry fields. This keeps the user's attention focused on the cells he is permitted to edit, and facilitates navigation among these cells.

When "Select locked cells" is disabled, the user can not see the source code contained in locked cells while viewing the spreadsheet Normal View. However, in Formula View, the user can see (but not edit) cell formulas. (Note that Formula View is toggled with the keyboard shortcut ctrl + ` where "`" is the grave accent.) To insure that the user can not see the source code, disable "Select locked cells" and insure that Locked cells are also Hidden (section 8.1).

Note that if the worksheet is Protected with "Format cells" enabled, the user can access the Format Cells dialog box, which is the place where cells are Locked and Hidden. A devious user might think he could change the format of a cell to alter its Cell Protection Format to unLock or unHide it. Fortunately, when Sheet Protection is enabled, the \Protection\ tab within the Cell Formatting dialog box is not available. (Drat, foiled again!)

The programmer can secure Sheet Protection using a password (section 8.5).

**8.3. Hiding Worksheets: Make Source Code Unavailable to Users**

An effective technique for securing source code is to place it on worksheets that are Hidden and hence unavailable to the user. To Hide or unHide a worksheet, use Format\Sheet\Hide\ and Format\Sheet\Unhide\ or *right-click worksheet tab\Hide\*. The user can only learn about a hidden worksheet by using Format\Sheet\Unhide\. The programmer can use Workbook Protection (section 8.4) to prevent the user from unHiding any Hidden worksheets.

Although Visual Basic for Applications (VBA) which is built into Excel, and its big brother Visual Basic (a programming language often used for writing traditional applications) are outside the scope of this paper, it is worth mentioning that the programmer can use VBA to hide a worksheet. VBA even has the capability to make a worksheet "very hidden". A "very hidden" worksheet cannot be discovered or unhidden using Format\Sheet\Unhide\ in Excel and can only be unHidden from within VBA.

**8.4. Workbook Protection: Control Worksheets**

Workbook Protection controls whether a user can modify the worksheets or resize windows in Excel. Workbook Protection is controlled in the Protect Workbook dialog box, accessed via Tools\Protection\Protect Workbook\ or *Review\Changes\Protect Workbook\*.

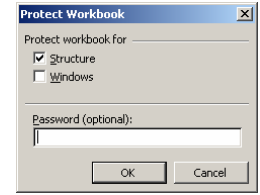

Enabling Structure means that the user can not move, delete, hide, unhide, rename, or insert new worksheets. Enabling Windows means that the user can not resize or move Windows. The ability to prevent users from unHiding worksheets is particularly important if the programmer has chosen to Hide worksheets.

The programmer can secure Workbook Protection settings using a password (section 8.5).

**8.5. Excel Passwords: Smart Practice, But Not Robust Protection**

The above-mentioned features to limit access to source code can be secured using a "workbook element" password. I recommend their use. Passwords will stymie the casual user, and likely provide adequate protection for most situations. However, a determined user can evade the password controls.

Excel has two types of passwords. The password to open a file uses high-quality encryption and is difficult to crack. Unfortunately, the passwords to protect workbook elements use very poor encryption. [McGimpsey 2004] has a clear explanation of the inherent weaknesses of these workbook element passwords prior to Excel 2007. Google on "crack Excel password" and you will see a mini-industry of nefarious tools. It is easy to obtain software that will quickly evade a workbook element password, and sometimes crack an open-file password.

Excel workbook element passwords should be viewed as a smart practice, providing sufficient source code protection for common situations. However, a moderately energetic user will be able to circumvent Excel workbook element passwords and gain access to the source code. This should be a concern for spreadsheets containing sensitive data, or requiring high source code integrity, such as those subject to Sarbanes-Oxley compliance.

**9. PROTECTING SOURCE CODE IN MICROSOFT EXCEL**

Excel has a set of source code protection features that work reasonably well provided they are carefully applied, and users do not take the effort to crack the workbook element passwords. Excel protections may be under-utilized, perhaps because they are so cumbersome to use. Excel source code protection recommendations are summarized below.

**9.1. Recommendations for Source Code Control in Microsoft Excel**

Spreadsheet application programmers can safeguard Excel source code by taking all of the following steps before deployment of a spreadsheet application to users.

Cell Protection Formatting
The default cell protection is Locked and unHidden. To obtain the strongest protection, the programmer should change the Protection Format of *all* cells in the spreadsheet:

- Empty cells:                          Format as Locked and Hidden.
- Cells containing source code:         Format as Locked and Hidden.
- Cells containing data entry fields:   Format as unLocked and unHidden.

Sheet Protection
The programmer should protect the worksheet. This is required to activate the Cell Protection Formats. When protecting the worksheet, enable the minimum necessary access. Under "Allow all users of this worksheet to:" the programmer should:

- enable (check)       "Select unlocked cells"
- disable (uncheck)    all other options

Hide Worksheets
Hide any worksheets that do not contain inputs or outputs intended for the user.

Workbook Protection
The programmer should protect the workbook. Enable "Structure".

Passwords
Use password protection for the elements that allow it, which are Sheet Protection and Workbook Protection. These passwords can be circumvented using readily available software.

**9.2. Make Copying of Hidden Cells Difficult**

There are many choices and options when protecting Excel source code. In some circumstances, it is possible for a user to evade protections. It works like this. The user selects two input cells while holding down the shift key, which selects a range of cells. This range includes the two input cells, plus all cells in the resulting rectangular range including any protected cells. With certain sets of protection settings, a copy-and-paste will reveal the source code. Section 9.1 is so detailed and specific in part to prevent copy-and-paste evasions.

It is smart practice to design the application user interface to prevent the possibility of this evasion. Place all inputs are in a common location, with no source code in the "rectangle" between any selectable input cells. One way to do this is to have a distinct input module. Better yet, place all cell formulas on a separate worksheet and Hide it.

**9.3. Excel Provides "Pretty Good Protection", or Lock the Door When Leaving Home**

Although cumbersome to use, Excel provides what might be called "pretty good protection" of spreadsheet source code.

Using the Excel protections with workbook element passwords is similar to locking the door to your home while hiding a key under the flowerpot. The door is locked, but an enterprising person can find the key and gain entry. This is fine if you live in a good neighborhood or have little that is worth stealing, but is unwise if you live in a rough part of town or have an overflowing treasure chest in the front hall.

Failure to use the workbook element passwords is like leaving your front door closed but unlocked. Failure to use the protections at all is like leaving your front door swinging in the wind: it's an invitation to improper entry.

The implications are clear: *User access to the source code is not an attribute of the Excel spreadsheet, but is a choice by the programmer.* Accepting the default of full access is a choice, not a given. The programmer could choose to protect the code.

**10. FEATURES REQUIRED IN LOCAL SPREADSHEETS TO SUPPORT APPLICATION DEVELOPMENT: THE CASE OF MICROSOFT EXCEL**

In aggregate, Excel can be considered a "pretty good" language for application development in situations where users are generally compliant. However, it could be much better.

Section 5 discusses six spreadsheet language features to support application programming. The essential element of source code control (feature 1), including locking all cells to prevent user customization (feature 2) is discussed in sections 8 and 9. Unfortunately, because a user can easily link his own spreadsheet into an Excel application, it seems inherently difficult to prevent linking (feature 3). Preventing users sharing the application (feature 4) also seems difficult to do through technical means, although strong managerial policies might be helpful.

The Excel protections are cumbersome and confusing to use. To obtain full protection requires the programmer to use four different tools, each with multiple combinations of settings and questionable defaults, plus two different passwords. Excel needs a purposefully designed source code protection tool (feature 5). This would include a new user interface that makes it simple to enable all protections necessary for complete protection, safeguarded by a single password. Any such facility should be carefully tested for evasions.

Passwords for source code control (feature 6) in local spreadsheets are, at least in theory, problematic. Securing the source code in a local spreadsheet is in some sense more difficult than securing SaaS source code because the password hash or some other "secret" must be resident in the spreadsheet file. Even an excellent encryption scheme is (at least "in theory") vulnerable to the depredations of the crafty people who have cracked other encryption schemes, including expensively-developed protections used for DVDs.

Despite this theoretical difficulty, it is likely that the shoddy encryption system currently used for Excel workbook element passwords [McGimpsey 2004] could be greatly improved. Increasing the key space from its current 194,560 to something much larger (e.g., 128 bits) would slow password cracking software, perhaps to the point of impracticability.

## 11. CONCLUSIONS AND FURTHER RESEARCH

This paper explores a number of ideas around spreadsheets. The main points are 1) spreadsheets are a computer programming language used for application development, 2) that this is a good thing, and 3) it would be a better thing if spreadsheet languages were designed to provide six application development features. The paper explores the status and prospects of application development features in a local spreadsheet (Microsoft Excel) and a software-as-a-service spreadsheet (Google Spreadsheet).

It is desirable that designers of spreadsheet computer programming languages incorporate application development features into their products. This is different from "bolting-on" poorly integrated protection features, as seems to be the case with Excel. I propose design changes to a SaaS and a local spreadsheet. Although the proposed design changes might appear to be simple, making changes to complex software products is never easy and might not be feasible. It is better to incorporate design-for-application-development in the early stage of creating a spreadsheet computer programming language.

Excel provides adequate protection of source code, provided the programmer chooses to use it, although the protections can be defeated by a determined user who applies password-cracking software. Version control will be improved by source code protection, but because users will be able to write their own spreadsheets that hook into the application, it will still be a challenge. Because Excel is a local spreadsheet, the problem of proliferation will persist.

Google Spreadsheet provides no protection of source code. However, the Software-as-a-Service spreadsheet delivery model has excellent potential for providing a strong application development platform, because it is possible (at least in principle) to provide the complete package of application development features. It is to be hoped that researchers or spreadsheet publishers will implement these features so that we can experiment with them and learn more.

Spreadsheet "lock box" approaches merit attention. For example, a service provided by Risk Integrated (www.riskintegrated.com) accepts an Excel file for deposit in a secure location, and runs it for clients who access inputs and outputs over the internet [Jafry et al 2006, 2007]. It provides all six spreadsheet application development features (section 5), plus change logging, audit trails, Monte Carlo simulation, and more.

A number of research opportunities regarding spreadsheet applications are emerging. What programming and software engineering practices make sense for building spreadsheet applications? For example, is there an alternative to the messy and risky nested IF constructs used for contingent logic? Can spreadsheet files be treated like subroutines, and sensibly linked into a full application, as is done with compiled software?

Handling data for a spreadsheet application has a host of issues. How can a programmer insure that users enter meaningful data? How can a manager insure that multiple people use the same data? How can an analyst efficiently run multiple large data sets through a spreadsheet application? The possibility of data sets with different numbers of rows and columns is but one complicating factor. These problems might soluble through a combination spreadsheet language features, spreadsheet programming practice, and user practice.

There are over a dozen spreadsheet languages, and it would be useful to explore and compare the application development features in each of them.

This paper focuses on purposeful applications. What can be done regarding *de facto* applications? Is there a migration path to something more robust? Can we enhance the quality (or slow the decay over time) of accidental spreadsheet legacy applications?

We would greatly benefit from empirical research on all aspects of spreadsheet application development, usage, and evolution. The vast world of spreadsheet users and developers will provide many gratifying surprises and opportunities to those who seek.

**ACKNOWLEDGEMENTS**

I am grateful to JP Allen, Steve Alter, Greg Benson, Mike Middleton, and Terence Parr at the University of San Francisco for their insights. Two referees provided valuable inspiration. Any errors of omission or commission are the sole responsibility of the author.

**REFERENCES**

Bewig, P. L. (2005), "How do you know your spreadsheet is right? Principles, Techniques and Practice of Spreadsheet Style", available at http://www.eusprig.org/, accessed February 27, 2007.

Gardner, D. W. (2006), "Spreadsheet Management Business Expected To Boom Fixing Old Error-Ridden Spreadsheets", Information Week, December 13, http://www.informationweek.com/showArticle.jhtml?articleID=196603875&cid=Answers, accessed February 28, 2007.

Grossman, T. A., V. Mehrotra, Ö. Özlük (2006), "Lessons from Mission Critical Spreadsheets", University of San Francisco School of Business and Management Working Paper.

Jafry, Y., F. Sidoroff, R. Chi (2006), "A computational framework for the near-elimination of spreadsheet risk", European Spreadsheet Risks Interest Group Proceedings, Cambridge.

Jafry, Y. C. Marrison and U. Umkehrer-Neudeck (2007), "Spreadsheet-based Monte Carlo simulation of real-estate credit risk", accepted by *Interfaces*.

Laudon, K. C. and J. P. Laudon (2006) Management Information Systems: Managing the Digital Firm, 9th Edition, Upper Saddle River, NJ: Prentice Hall.

McGimpsey, N. (2004), "Removing Internal XL Passwords", http://www.mcgimpsey.com/excel/removepwords.html, accessed March 4, 2007.

O'Beirne, P. (2005), "Spreadsheet Check and Control", Systems Publishing.

Powell, S. G. and K. R. Baker (2007), "Management science: The art of modeling with spreadsheets, second edition", John Wiley & Sons, Inc.

Read, N. and J. Batson (1999), "Spreadsheet modeling best practice", available at http://www.eusprig.org/ accessed February 27, 2007.

Wikipedia (2007), "Software as a Service", http://en.wikipedia.org/wiki/SaaS, accessed May 23, 2007.

Blank Page